\documentclass{emulateapj}

\usepackage{amsmath}

\begin{document}

\title{Pico: Parameters for the Impatient Cosmologist}

\author{William A.~Fendt\altaffilmark{1} and
        Benjamin D.~Wandelt\altaffilmark{1,2,3,4}}

\altaffiltext{1}{Department of Physics, UIUC, 1110 W Green Street, 
             Urbana, IL 61801; fendt@uiuc.edu}
\altaffiltext{2}{Department of Astronomy, UIUC, 1002 W Green
             Street, Urbana, IL 61801; bwandelt@uiuc.edu}
\altaffiltext{3}{Center for Advanced Studies, UIUC, 912 W Illinois 
             Street, Urbana, IL 61801}
\altaffiltext{4}{Benjamin D.~Wandelt is a Center for Advanced Studies 
             Beckman Fellow}


\begin{abstract}
We present a fast, accurate, robust and flexible method of accelerating parameter
estimation.
This algorithm, called Pico, can compute the CMB power spectrum and matter transfer 
function as well as any computationally expensive likelihoods in a few milliseconds.
By removing these bottlenecks from parameter estimation codes,
Pico decreases their computational time by $1$ or $2$ orders of magnitude.
Pico has several important properties.  
First, it is extremely fast and accurate over a large volume of parameter space. 
Furthermore, its accuracy can continue to be improved by using a larger training set. 
This method is generalizable to an arbitrary number of cosmological parameters 
and to any range of $\ell$-values in multipole space.
Pico is approximately $3000$ times faster than CAMB for flat models, and
approximately $2000$ times faster then the WMAP $3$ year likelihood code.
In this paper, we demonstrate that using Pico to compute power spectra and
likelihoods produces parameter posteriors that
are very similar to those using CAMB and the official WMAP3 code, 
but in only a fraction of the time.
Pico and an interface to CosmoMC are made publicly available 
at \verb+www.astro.uiuc.edu/~bwandelt/pico/+.
\end{abstract}

\keywords{cosmic microwave background --- cosmology: observations ---
          methods: numerical}


\section{Introduction}\label{intro}
With WMAP's second data release \citep{Hinshaw:2006ia,Page:2006hz,Spergel:2006hy}, 
there is a wealth of new CMB data available to further constrain the
cosmological parameters.  
The major computational burden in parameter estimation remains 
the calculation of the theoretical power spectrum for a large number 
of cosmological models as well as the likelihood based on these spectra. 
Generally the power spectrum is computed with codes such as CMBfast 
\citep{Seljak:1996is} or CAMB \citep{Lewis:1999bs}, 
which evolve the Boltzmann equation using a line of sight integration approach.
While this provides a $1$ or $2$ order of magnitude decrease in
the computation time over the full Boltzmann codes, power spectrum 
calculations remain a bottleneck of parameter estimation.
Other software such as CMBwarp \citep{Jimenez:2004ct} and DASh 
\citep{Kaplinghat:2002mh} have found ways to improve the efficiency of 
power spectrum calculations at the cost of a loss of accuracy against 
the full Boltzmann codes and/or placing restrictions on the parameters 
which are available as input.
In particular, CMBwarp builds upon the method introduced in \citep{Kosowsky:2002zt}, 
where a new set of nearly uncorrelated ``physical'' parameters were defined 
which have nearly independent effects on the power spectrum.
CMBwarp uses a modified polynomial fit whose coefficients are based on a 
fiducial model.
It allows rapid calculation of the temperature (TT), E-mode polarization 
(EE) and temperature-polarization (TE) cross power spectra.
CMBwarp, however, requires the use of specific cosmological parameters
and the accuracy of the computed power spectra quickly diminishes as 
one moves away from the fiducial model in parameter space.
Another code, CMBFit \citep{Sandvik:2003ii}, attempts to avoid the need
to compute the power spectrum by fitting the likelihood function.  
This idea is particularly important for the WMAP $3$ year data 
\citep{Hinshaw:2006ia,Page:2006hz,Spergel:2006hy}
whose likelihood is time consuming to compute.

In this paper we introduce Pico, a computational technique to
accelerate both power spectrum and likelihood computations.
This approach removes the $2$ major bottlenecks in parameter
estimation.
While in a similar spirit as CMBwarp, and providing a similar speedup 
over CMBfast and CAMB, Pico has several important advantages
over CMBwarp and DASh. 
First, it allows the calculation of power spectra from an arbitrary 
number of cosmological parameters and in any range of $\ell$-values
in multipole space.
Because of this flexibility, it is  easily incorporated into 
parameter estimation codes.
Secondly, Pico allows the simultaneous computation of all scalar, 
tensor and lensed power spectra as well as the transfer functions.  
Pico provides more than an order of magnitude increase in 
accuracy over CMBwarp, and about $2$ orders of magnitude increase in 
speed over DASh. 
Lastly, Pico is generic enough to allow the direct fitting of any likelihood 
functions. 
Due to the computational expense in computing the likelihood of certain experiments,
e.g. WMAP3, any power spectrum acceleration scheme will at most provide a speed up
of order $1$ to $10$ in parameter estimation.
However, using Pico to also compute the likelihood results in speed ups
of order $10$ to $100$.
As an additional bonus, using Pico to compute the likelihood directly provides 
more accurate results then using it to fit the power spectra and computing the 
likelihood from these approximate spectra.  
This is important for current and next generation all-sky CMB data.
Meanwhile, the power spectra computed by Pico are more than accurate enough for suborbital
experiments with smaller sky coverage and coarser $\ell$-resolution.

This paper is organized as follows.  
We examine the CPU and memory requirements of Pico in section \ref{cpu}.
Section \ref{results} presents several tests of the performance of Pico.
This includes comparisons of power spectra computed using Pico and CAMB
as well as results of parameter estimation runs using Pico to compute the 
power spectra and the WMAP3 likelihood.
In section \ref{concl} we summarize and discuss the future of Pico.
The details of the algorithm used by Pico are presented in the appendix.


\section{CPU and Memory Requirements}\label{cpu}

For a detailed description of our algorithm please refer to the appendix.
The quantities that determine the CPU and memory requirements of the 
algorithm are the number of clusters $n$, the number of cosmological
parameters $\mathcal{N}_{x}$, the number of $\ell$-values and 
compressed $\ell$-values $\mathcal{N}_{y}$ and $\mathcal{N}_{y}^{\prime}$
and the order of the regression polynomial $p$.
Each computation of the power spectra has (approximately) no dependence on the 
number of clusters, since we only need to determine which cluster the
input parameters are in. This is found after $n$  fast distance
calculations.  The power spectrum is then calculated using the polynomial
in the cluster.  This takes $2\mathcal{N}\mathcal{N}_{y}^{\prime}$ computations where
$\mathcal{N}$, the number of polynomial coefficients, is given by 
\begin{equation*}
   \mathcal{N} = \frac{\left(\mathcal{N}_{x}+p\right)!}{\mathcal{N}_{x}!p!}
       \sim \mathcal{O}\left(\left(\frac{\mathcal{N}_{x}}{p}\right)^{p}\right)
       \;\;\; \mbox{for } \mathcal{N}_{x} \gg p \gg 0.
\end{equation*}
After evaluating the polynomial, another $\mathcal{N}_{y}\mathcal{N}_{y}^{\prime}$
calculations are needed to uncompress the spectrum.
It is thus possible to calculate the power spectrum with very few
computations.  
For the $7$ parameter case we examine in section \ref{test1}
calculation of the power spectra takes approximately $3$ milliseconds
on a 2 GHz Intel Pentium M processor.
This is roughly $3000$ times faster then CAMB for flat models and
$15000$ times faster for nonflat models.  
For parameter estimation codes such as CosmoMC \citep{Lewis:2002ah}, this speedup
is significantly more than is necessary since evaluation of the likelihood quickly
becomes the new bottleneck.  However, as we have noted, Pico removes this bottleneck 
as well by fitting the (computationally intensive) likelihoods.
Furthermore, other techniques such as Gibbs sampling \citep{Wandelt:2003uk,Chu:2004zp}, 
which have quicker likelihood evaluations, continue to benefit from a significant 
speedup in computing the power spectrum.

The main memory use of the algorithm is in holding the 
$n \mathcal{N} \mathcal{N}_{y}^{\prime}$ regression coefficients.
For the example we present in the next section using $4^{\mathrm{th}}$
order polynomials in $7$ parameters ($\mathcal{N}=330$) over $100$ clusters
and $60$ compressed $\ell$-values, this is approximately $15$ MB of information.  
If fitting of the scalar and tensor modes  out to $\ell=3000$
as well as the transfer function is included this number could 
increase by an order of magnitude. 
Even this can be accommodated by any modern personal computer.


\section{Results}\label{results}
In this section we provide several tests of the performance of Pico both in its 
ability to compute power spectra and in the results of parameter 
estimation runs.  The first $2$ tests compare the power spectra
computed using Pico with those using CAMB for $7$ and $9$ parameter models.
Next we compare the results of a parameter estimation run using Pico and CAMB to
compute the power spectrum and the $1^{\mathrm{st}}$ year WMAP code 
\citep{Bennett:2003bz} to compute the likelihood.
Lastly, we compare parameter estimation runs using Pico to compute the likelihood with 
runs using CAMB and the official WMAP $3$ year likelihood code.
In all cases Pico produces parameter posteriors that are in good agreement with CAMB,
both for the larger parameter space allowed by WMAP1 and the higher precision 
required by WMAP3.

\subsection{Power Spectrum Calculation for $7$ Parameter Models}\label{test1}
Here we compare the performance of Pico with CAMB and CMBwarp.
To generate our test set we begin with a converged Markov chain Monte Carlo 
run consisting of $\sim 60000$ cosmological models based on WMAP $1^{\mathrm{st}}$
year \citep{Bennett:2003bz} and other CMB data, including 
CBI \citep{Padin:2001df}, BOOMERANG \citep{Ruhl:2002cz}, 
ACBAR \citep{Kuo:2002ua}, VSA \citep{Grainge:2002da},
MAXIMA \cite{Hanany:2000qf}, DASI \citep{Halverson:2001yy} and
TOCO \citep{Miller:1999qz}.
The varied parameters are the baryon density $\Omega_{\mathrm{b}}$, the cold dark 
matter density $\Omega_{\mathrm{cdm}}$, the dark energy density 
$\Omega_{\mathrm{\Lambda}}$, Hubble's constant $H_\mathrm{0}$, the scalar 
spectral index $n_\mathrm{s}$, the optical 
depth since reionization $\tau$, and the normalization of the power spectra 
$\mathcal{A}_{\mathrm{s}}$. 
Next we convert each point in this space to the physical parameters introduced 
by \citet{Jimenez:2004ct} and \citet{Kosowsky:2002zt}. 
In the physical parameter space there is significantly less correlation
in the set of points.  
We then calculate the mean and variance of this set, and use it to generate 
an $8$-dimensional box in physical parameter space whose sides are of length 
$3\sigma$ in each direction.
Our test set consists of $10^4$ models sampled {\it uniformly} from this box.  
That is, we in no way bias the test set by weighting points in parameter space 
based on their likelihoods. 
The physical parameters are converted back to cosmological parameters and used 
to run CAMB to give the scalar TT, TE and EE power spectra.
This set of $10^4$ models and their corresponding power spectra form our test set.
 
\begin{figure*}
   \begin{center}
      \resizebox{150mm}{!}{\includegraphics{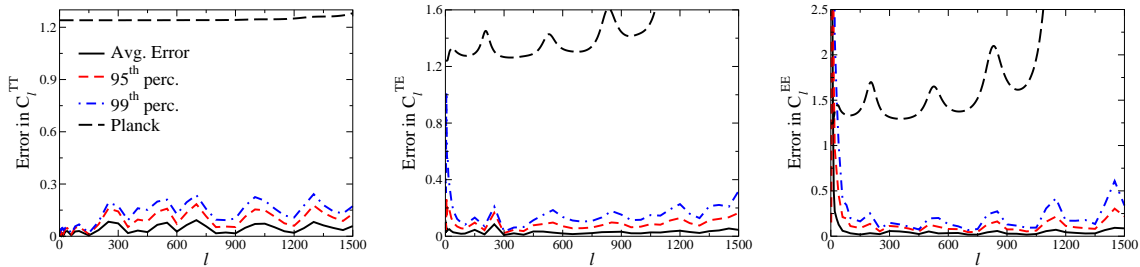}}
      \caption{The above plots compare the performance of Pico
               with CAMB for $7$ parameter models. The three lines 
               denote the average error and the $95^{\mathrm{th}}$ 
               and $99^{\mathrm{th}}$ percentiles over the $10^{4}$
               models in the test set. 
               The dashed black line is the expected uncertainty from 
               Planck data assuming $f_{sky}=0.65$.
               The error is 
               plotted in units of the cosmic standard deviation.}
      \label{result-pics}
   \end{center}
\end{figure*}

The performance of Pico is shown in Figure (\ref{result-pics}). 
In this example, we ran Pico with $4^{\mathrm{th}}$ order polynomials over $100$ 
clusters.
We have plotted the error in units of the cosmic standard deviation as a function 
of the multipole $\ell$-value for the TT, TE and EE power spectra.  
The error in a single computed spectrum is defined as
\begin{equation*}
   \Delta_{\ell} = \frac{|C_{\ell} - C_{\ell}^{\mathrm{CAMB}}|}{\sigma_{\ell}^{CV}},
\end{equation*}
where $\sigma_{\ell}^{CV}$ is the cosmic standard deviation.
The three lines denote the average of this error over the test set and the error 
which bounds $95\%$ and $99\%$ of the test set (the $95^{\mathrm{th}}$ and
$99^{\mathrm{th}}$ percentiles).  
The dashed black line in each plot denotes the expected
uncertainty in data from the Planck satellite mission.
Here we have assumed $65\%$ of the sky will remain uncontaminated by foregrounds, and
we have combined the $3$ frequency bands from the LFI and the $3$ lowest frequency
bands from the HFI according to the method described by \citet{Zaldarriaga:1997ch}
and \citet{Kinney:1998md}.

For $99\%$ of the models in our test set, Pico is able to calculate 
the TT spectrum with an error less then $0.3\sigma^{CV}$, the TE spectrum with 
an error less then $0.4\sigma^{CV}$ and the EE spectrum to better then 
$0.7\sigma^{CV}$ for $\ell$ out to $1500$. 
For the TE and EE spectra this excludes very low $\ell$ where the magnitudes of the
power spectra and cosmic variance become small.
This is better than what will be achievable from even the Planck satellite mission.
We note that the points with the largest error bars are near the edges of our
training set and correspond to models that are highly disfavored even by CMB
data alone.

\begin{figure*}
   \begin{center}
      \resizebox{150mm}{!}{\includegraphics{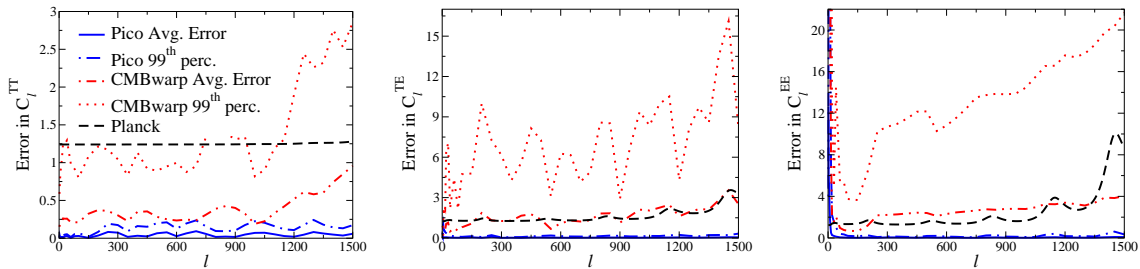}}
      \caption{The above plots compare the performance of Pico
               with CMBwarp. The four lines denote the average error 
               and $99^{\mathrm{th}}$ percentile for CMBwarp and the 
               average error and $99^{\mathrm{th}}$ percentile using
               Pico. 
               The dashed black line is the expected uncertainty from 
               Planck data.
               The error is plotted in units of the cosmic 
               standard deviation.} 
      \label{cmbwarp-pics}
   \end{center}
\end{figure*}

In Figure (\ref{cmbwarp-pics}), we have plotted the performance of 
CMBwarp against CAMB over the same test set.  
The four lines denote the average and $99^{\mathrm{th}}$ percentile from 
CMBwarp as well as the average and  $99^{\mathrm{th}}$ percentile using our code. 
We note that Pico is significantly more accurate over all $\ell$
for the $3$ power spectra.
Pico gives more than an order of magnitude increase in accuracy over CMBwarp, 
while providing a similar decrease in the time required to compute a power 
spectrum as compared with CAMB. 

\subsection{Power Spectrum Calculation for $9$ Parameter Models}\label{test2}
\begin{figure*}
   \begin{center}
      \resizebox{150mm}{!}{\includegraphics{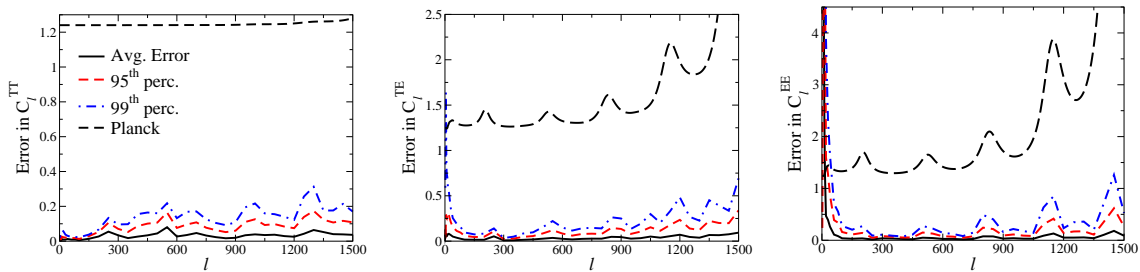}}
      \caption{The above plots compare the performance of Pico
               with CAMB for $9$ parameter models. The three lines 
               denote the average error and 
               the $95^{\mathrm{th}}$ and $99^{\mathrm{th}}$ 
               percentiles. 
               The dashed black line is the
               expected uncertainty from Planck data.
               The error is plotted in units of the
               cosmic standard deviation.}
      \label{result-ep-pics}
   \end{center}
\end{figure*}

As a second test of Pico, we calculate the scalar TT, TE and
EE spectra as a function of $9$ parameters.
The training set was formed using the $7$ parameters as described 
in section \ref{test1}, 
in addition to the dark energy equation of state and the
running of the scalar spectral index. The new parameters were
 drawn uniformly from the intervals
$\left[-1,-0.78\right]$ and $\left[-0.085,0\right]$ respectively.
Figure (\ref{result-ep-pics}) shows the performance compared with CAMB.
In this example Pico was run with $4^{\mathrm{th}}$ order polynomials over
$100$ clusters.  
While even at this level the accuracy is better then what will be achievable 
with Planck, one could continue to decrease the
error by using a larger training set to allow the use of more clusters.

\subsection{Parameter Posteriors using Pico to compute Power Spectra}\label{test3}
We have incorporated Pico into the publicly available parameter estimation 
code CosmoMC \citep{Lewis:2002ah}.  
The interface allows CosmoMC to use Pico to compute the theoretical power spectrum 
and transfer function as well as the WMAP3 likelihood whenever 
the parameters are within the range over which Pico's regression coefficients are 
defined.  
For parameters outside this range, CosmoMC will continue to use CAMB to compute 
the power spectrum or the WMAP3 code to compute the likelihood.

\begin{figure*}
   \begin{center}
      \resizebox{150mm}{!}{\includegraphics{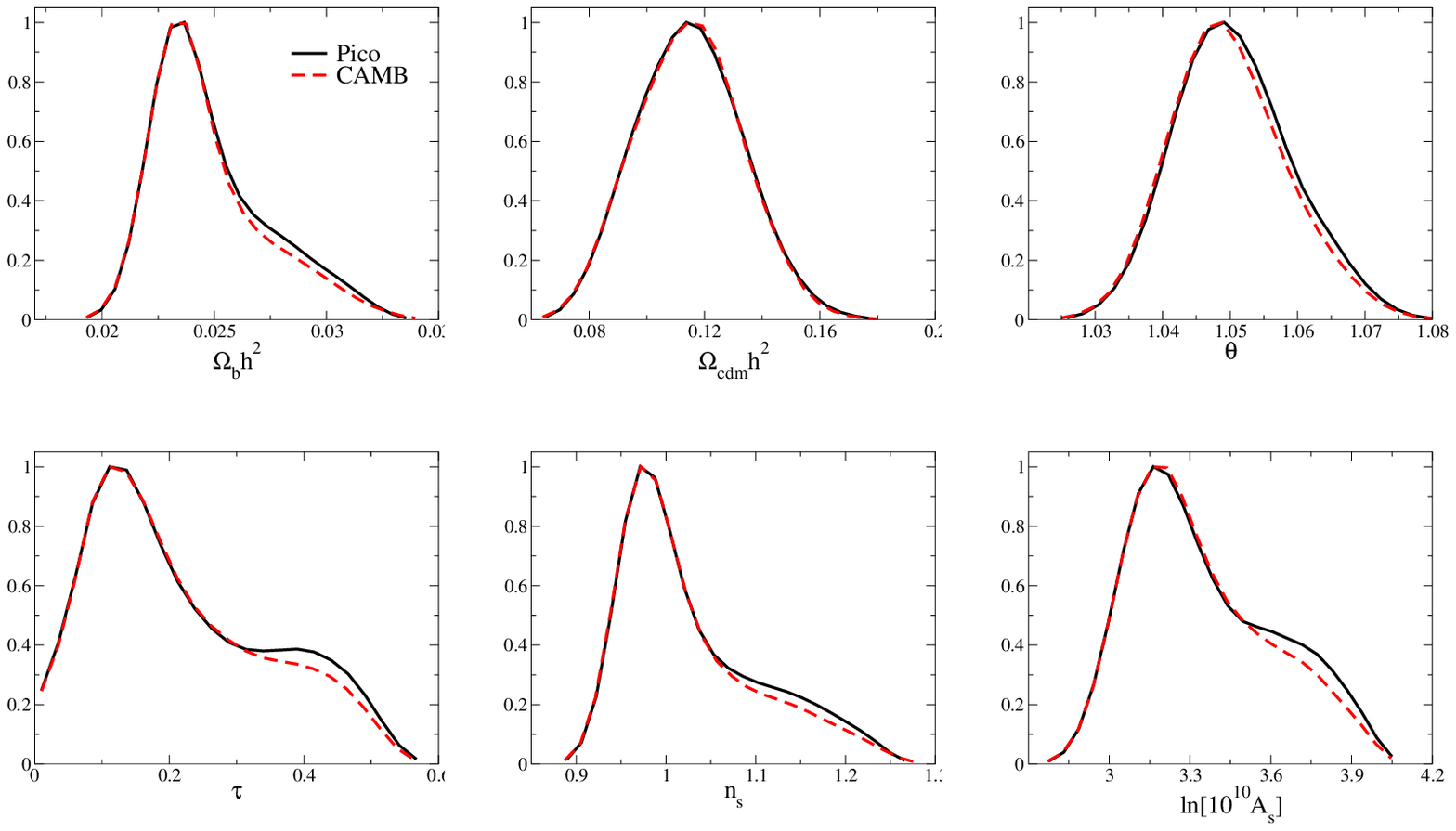}}
      \caption{The $1$-D posterior constraints on the cosmological parameters using 
               runs of CosmoMC with Pico and CAMB. 
               The likelihoods were computed using the $1^{\mathrm{st}}$
               year WMAP data and likelihood code based on the Pico power spectra
               (black, solid line) and the CAMB power spectra (red, dashed line).
               Using Pico decreased the time to compute the power spectra by a 
               factor of $3000$, and the overall computational time by a factor of $10$, 
               while providing very similar posteriors.}
      \label{result-post}
   \end{center}
\end{figure*}

\begin{figure*}
   \begin{center}
      \resizebox{150mm}{!}{\includegraphics{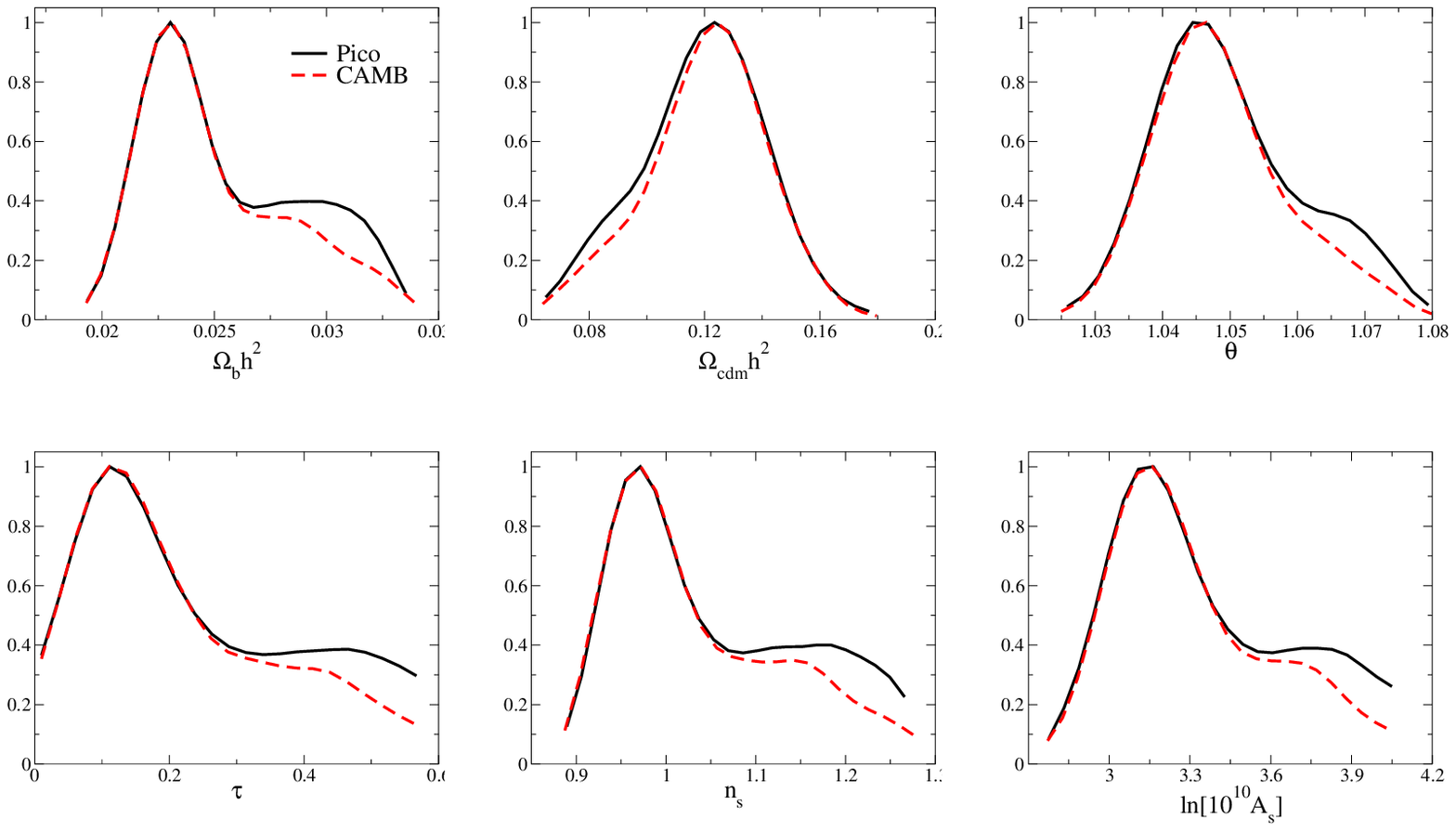}}
      \caption{The mean likelihoods over the cosmological parameters using 
               runs of CosmoMC with Pico and CAMB. 
               The likelihoods were computed using the $1^{\mathrm{st}}$
               year WMAP data and likelihood code based on the Pico power spectra
               (black, solid line) and the CAMB power spectra (red, dashed line).
               Using Pico decreased the overall
               computational time by a factor of $10$, while accurately fitting the
               peaks of the mean likelihoods. 
               The error in the tails is due to correlated error in the Pico computed 
               power spectra.}
      \label{result-meanlikes}
   \end{center}
\end{figure*}

In this section we compare the posteriors over the parameters computed by
CosmoMC while using CAMB or Pico to compute the theoretical power spectrum.
The likelihoods were computed using the WMAP $1^{st}$ year data 
and likelihood function.
\citep{Verde:2003ey,Hinshaw:2003ex,Kogut:2003et}.
For this test we choose flat models and varied $6$ parameters:
$\Omega_{\mathrm{b}}h^2$, $\Omega_{\mathrm{cdm}}h^2$, $\theta$,
$\tau$, $n_{\mathrm{s}}$ and the power spectrum amplitude $\mathcal{A}_{\mathrm{s}}$.
We ran CosmoMC using Pico for $500,000$ steps. 
CAMB was needed for less than $1\%$ of the models.
This took approximately 15 hours.
The posterior and mean likelihood over each parameter is shown in figures
(\ref{result-post}) and (\ref{result-meanlikes}) respectively.
We have also plotted the posterior and mean likelihood 
from a 500,000 step run of CosmoMC using only CAMB, which took 
approximately 160 hours.
The posteriors agree quite well, especially near the peaks.  
In every parameter except $\tau$, the mean of the posteriors differ by less 
then $0.7\%$.
For $\tau$, which is poorly constrained by this data set, the mean of the posteriors
differ by $3.7\%$. 
The errors in the likelihood evaluations from Pico are more apparent 
in Figure (\ref{result-meanlikes}) as the mean likelihood over the posterior
depends on the square of the likelihood.  Also, the likelihood is very 
sensitive to any correlated errors in the approximate power spectra computed by Pico.
As will be shown in the following section, this problem is solved by using Pico
to directly compute the likelihood.
Even here, however, Pico agrees quite well with CAMB around the peak of the 
mean likelihood.

\begin{figure*}
   \begin{center}
      \resizebox{70mm}{!}{\includegraphics{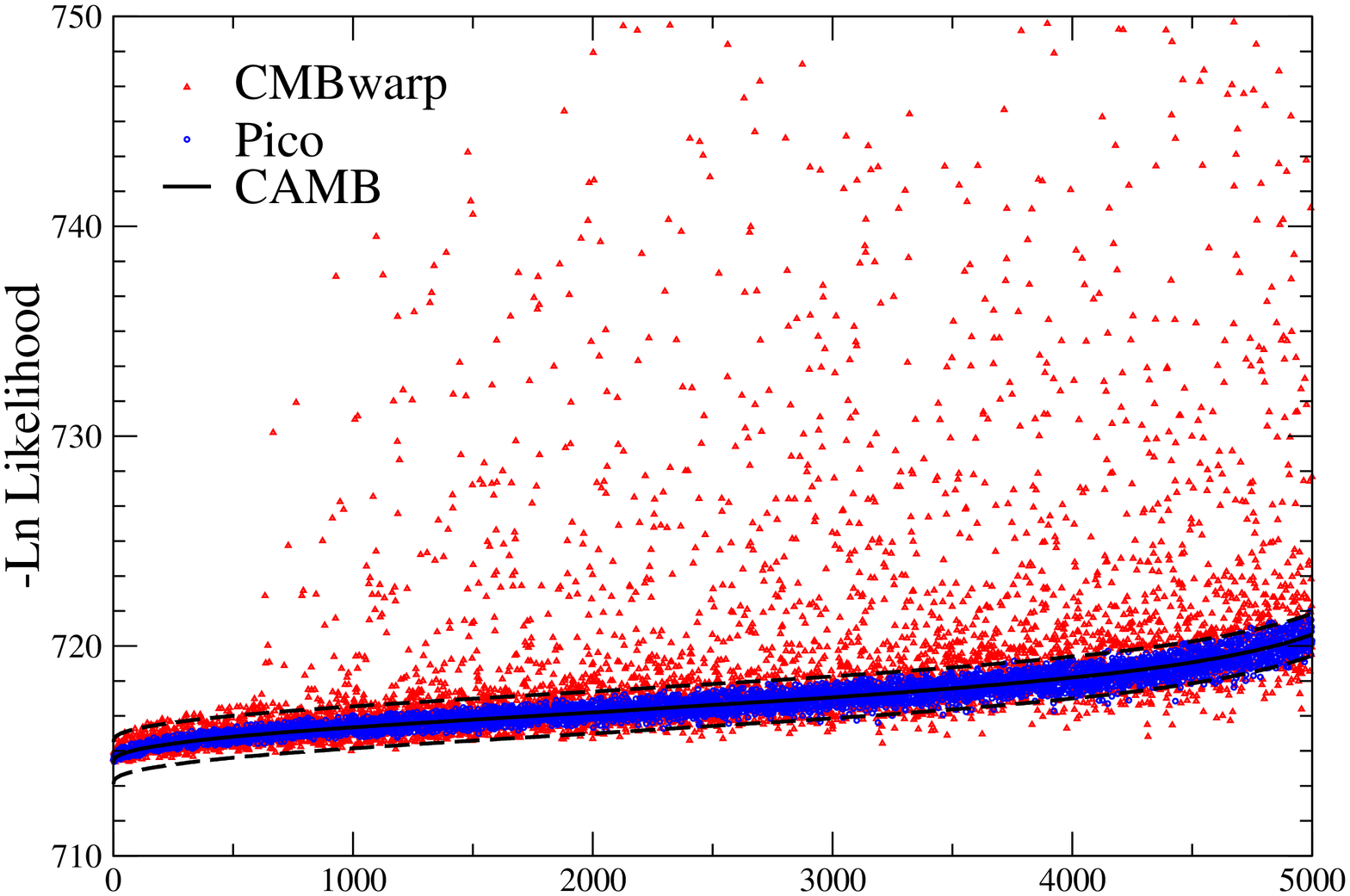}}
      \caption{A comparison of direct computation of the likelihood for a chain
               of models. The thin line is the value of the log likelihood using 
               power spectra from CAMB.
               The (blue) circles and (red) triangles are the values of the log
               likelihoods computed using power spectra from Pico and CMBwarp 
               respectively.
               The dashed black lines are $\pm1$ of the log likelihood using CAMB.
               Pico agrees within $1$ log likelihood over $2$ decades
               of likelihood values.}
      \label{likes_comp}
   \end{center}
\end{figure*}

In Figure (\ref{likes_comp}), we directly compare the accuracy of the likelihoods
computed using power spectra from Pico and CMBwarp with CAMB.
Using a uniformly sampled subset of the MCMC chain discussed in the 
previous paragraph, we first ordered the points by likelihood (black line). 
Next we recomputed the likelihood using power spectra from Pico (blue circles) and 
CMBwarp (red triangles) at each point.
We see that the error in Pico is less then unity over two decades in likelihood.  
This is a significant improvement over CMBwarp.
The dotted black lines are plus and minus one of the actual value of the log likelihood.

\subsection{Parameter Posteriors using Pico to compute the Likelihood}\label{test4}
\begin{figure*}
   \begin{center}
      \resizebox{150mm}{!}{\includegraphics{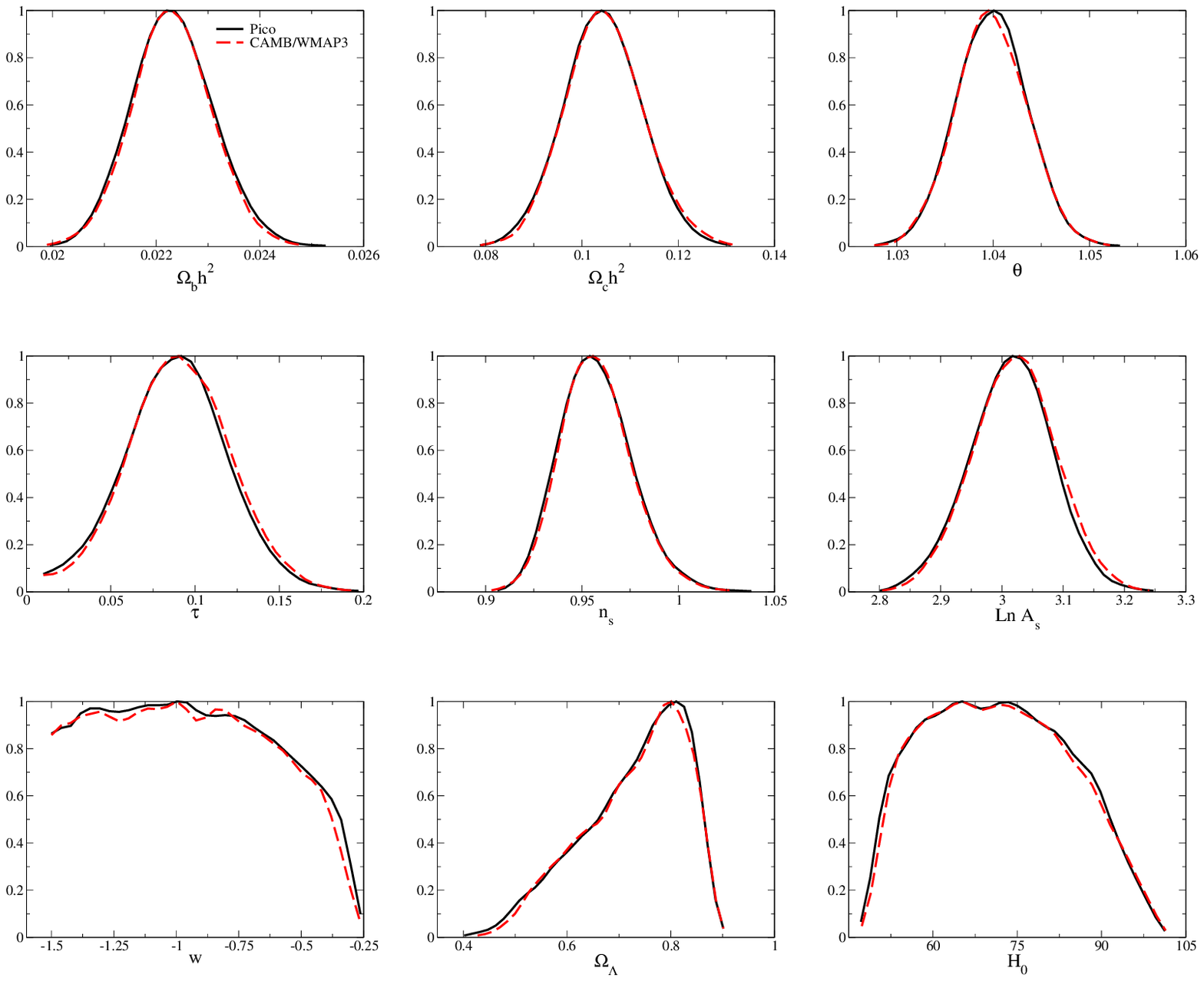}}
      \caption{The posteriors for $7$ parameter flat models using the WMAP 
               $3$ year data.
               The solid (black) line denotes the posterior using Pico to compute 
               both the power spectrum and the likelihood, while the dashed (red) 
               line is the posterior using CAMB and the WMAP $3$ year likelihood.
               The Hubble constant $H_{0}$ and the dark energy density 
               $\Omega_{\Lambda}$ are derived from the other $7$ parameters. 
               Using Pico to compute the likelihood and power spectra provides 
               a factor of $30$ increase in the speed of the parameter estimation 
               code. The improvement over Figure \ref{result-post} is the result 
               of fitting the likelihood directly with Pico.}
      \label{post2}
   \end{center}
\end{figure*}

As a final test, we ran CosmoMC using Pico to compute the WMAP3 likelihood.
Figure (\ref{post2}) compares the posteriors of this run with those using CAMB
and the official WMAP 3 year likelihood code.
The chains varied $7$ parameters; these included the $6$ parameters listed in 
section \ref{test3} as well as the dark energy equation of state $w$.  
The red, dashed line denotes the posterior using Pico to compute
the power spectrum \textit{and} the WMAP3 likelihood. 
The black, solid line denotes the posterior using CAMB and the 
official WMAP3 likelihood code. 
By fitting both the likelihood and the power spectrum Pico provides a factor of $30$
increase in speed.  Furthermore, with Pico, CosmoMC spends only about $1/5$ of its time 
computing the power spectrum and likelihood, demonstrating that Pico has successfully 
removed these two bottlenecks from the parameter estimation process.
Though it is not needed in this example, Pico also provides the transfer
function so that CAMB can compute the matter power spectrum and $\sigma_{8}$.

\section{Conclusion}\label{concl}
This paper provides a fast, accurate and robust method of calculating 
CMB power spectra and likelihood functions using local polynomial interpolation.  
A K-means clustering algorithm is
used to partition the cosmological parameter space into local regions.  
Over each region we approximate the CMB power spectra as a polynomial in 
the cosmological parameters.  
This method, which we have named Pico, provides several orders of magnitude
increase in speed over CAMB and the WMAP $3$ year likelihood code, 
while proving accurate enough for the analysis
of data from the current and next generation of cosmic microwave
background experiments.
The flexibility of our algorithm enables it to handle any reasonable number 
of cosmological parameters. 
It has been generalized to allow the fast computation 
of any observables relevant to a particular data set, e.g. the transfer functions and
the power spectrum of B-mode polarization anisotropies.
Even higher order correlation functions, such as the reduced bispectrum,
could be added.
Pico is able to compute accurate power spectra over a large volume of 
parameter space consistent with the WMAP data. 
Furthermore, Pico's performance will only improve as the volume of space it must 
fit and the uncertainties in the parameters shrink.

Pico is easily inserted into parameter estimation codes such as CosmoMC.
It can be used to compute the power spectra, transfer function and the WMAP3 
likelihood, resulting in a significant decrease in computational time.
In fact, when Pico is used, CosmoMC spends $80\%$ of its time on tasks 
other then computing the power spectra and likelihood. 
This time is spent generating random numbers, evaluating internal and derived 
parameters, etc.
We envision that CAMB will only be needed for nonstandard
cosmological models outside the scopes of our training sets.
While it is likely possible to further improve the accuracy of our code
by using less generic techniques, we have chosen to keep Pico as generic 
as possible to allow it to grow and adapt to the parameter estimation
tasks of the next generation of experiments.

We have made a Fortran 90 implementation of this algorithm publicly 
available.\footnote{ http://www.astro.uiuc.edu/\~{}bwandelt/pico/ }
Here the user will find regression coefficients to use the algorithm for various
parameter sets, as well as short and straight forward instructions for 
incorporating Pico into CosmoMC or using it as a front end for CAMB. 
The authors also welcome requests for regression coefficients for specific 
combinations and ranges of parameters.
Enabling Pico on a new parameter set simply involves running CAMB to 
generate a new training set.


\acknowledgements
We would like to thank Lloyd Knox, Antony Lewis and Max Tegmark for useful 
discussion. This work was partially funded by NSF grant AST 05-07676, 
by NASA contract JPL1236748, by the National Computational Science Alliance 
under AST300029N and by the University of Illinois. We utilized the
Teragrid (\texttt{www.teragrid.org}) 
Xeon cluster tungsten (\texttt{login-w.ncsa.teragrid.org}) and
Itanium 2 cluster mercury (\texttt{tg-login.ncsa.teragrid.org})
at NCSA.

\appendix

\section{Algorithm}\label{algorithm}
This appendix presents the basic algorithm Pico uses to calculate
the angular power spectra.
It consists of three major pieces, the compression of the training set
power spectra, the clustering of the training set cosmological parameters, 
and the calculation of the local regression polynomials. For clarity reasons,
we will discuss the latter of these pieces first.

\subsection{Polynomial Interpolation}\label{fitting}
Consider a training set of $N$ vectors of cosmological parameters 
$\mathbf{x}_j$, each of dimension $\mathcal{N}_{x}$ and their 
corresponding power spectrum $\mathbf{y}_j$, each of dimension $\mathcal{N}_{y}$.  
The number of cosmological parameters and power spectrum values is arbitrary.
In general, $\mathbf{y}$ can be constructed by concatenating all the 
scalar, tensor and lensed power spectra as well as the transfer functions
into a single vector living in $\mathbb{R}^{\mathcal{N}_{y}}$. 

Our goal is to interpolate the function $\mathbf{f}$ that maps the cosmological 
parameters $\mathbf{x}$ into their power spectra $\mathbf{y}$, i.e. 
$\mathbf{y}=\mathbf{f}\left(\mathbf{x}\right)$. 
This function is an $\mathcal{N}_{x}$ dimensional manifold that is 
naturally embedded in an $\left(\mathcal{N}_{x}+\mathcal{N}_{y}\right)$
dimensional Euclidean space.
Our method is to approximate this mapping using a polynomial in the
cosmological parameters.
The $k^{\mathrm{th}}$ component of $\mathbf{y}$ is then approximated as a
$p^{\mathrm{th}}$ order polynomial in the $\mathcal{N}_{x}$ cosmological
parameters:
\begin{equation*}
   y_{k} = \sum_{i_{1}\ge i_{2} \ge \cdots \ge i_{p}}^{\mathcal{N}_{x}}
           \alpha_{i_{1} i_{2} \cdots i_{p}} x_{i_{1}} x_{i_{2}} \cdots x_{i_{p}}.
\end{equation*}
The coefficients $\alpha_{i_{1} \cdots i_{p}}$ are chosen to minimize the 
squared error over the training set
\begin{equation*}
   R^{2} = \sum_{j=1}^{N} \left(\mathbf{y}\left(\mathbf{x}_{j}\right) - 
                                 \mathbf{y}_{j}\right)^{2}.
\end{equation*}
This leads to a regression matrix which can be inverted to find the polynomial 
coefficients. 

We have generalized this algorithm to include arbitrary fitting functions,  
for example Chebyshev or Legendre polynomials.
Our tests show that Pico performs at a similar level using these functions as
using standard polynomials.


\subsection{Clustering}\label{cluster}
The interpolation method described above fails to accurately model the
power spectra over the entire parameter space.  
To remedy this, we would like to fit polynomials on disjoint local regions of the 
full parameter space, limiting the variation in the power spectra over
the individual regions. 
While naively griding this large dimensional space would be computationally prohibitive,
 
clustering avoids the ``curse of dimensionality'' by using the points in the
training set to naturally divide the parameter space into smaller regions.  
A polynomial is used within each 
cluster to provide a local approximation of the power spectra within the cluster.  
It is then only necessary to ensure that each cluster has a sufficient
number of training set points to accurately calculate the regression coefficients.
We implement clustering using the K-means algorithm \citep{MacQueen:1967},
which we found adequate for our purposes.

Ideally all clusters encompass volumes of 
parameter space over which the power spectra vary roughly equally.
For example, we would like to take into account the fact that there is a roughly
equal variation of the power spectra from a change in the baryon density of
$\sim 0.01$ as from a change in the cold dark matter density of $\sim 0.1$. 
This is achieved by sphering the training set prior to clustering.
A sphered data set is defined to have a covariance matrix equal to the identity.
If $C_{xx}$ denotes the covariance matrix of the cosmological parameters that 
make up the training set, then the set is sphered by constructing the matrix
$MU$ such that
\begin{equation*}
   M U C_{xx} U^{T} M^{T} = M E_{xx} M = I.
\end{equation*}
Here $U$ is the orthogonal matrix that diagonalizes $C_{xx}$, $M$ is a diagonal matrix
whose entries are the inverse of the square root of the eigenvalues of $C_{xx}$,
and the diagonal matrix $E_{xx}$ contains the eigenvalues of $C_{xx}$.
The matrix $MU$ is used to map the training set into a new sphered basis.
In this basis the parameter space is clustered using the K-means algorithm 
and the standard Euclidean distance.
Since in the sphered space, the power spectra 
corresponding to the parameters will vary equally in all directions, the clusters 
will retain this desired property when mapped back to the unsphered basis.

\begin{figure*}
   \begin{center}
      \resizebox{100mm}{!}{\includegraphics{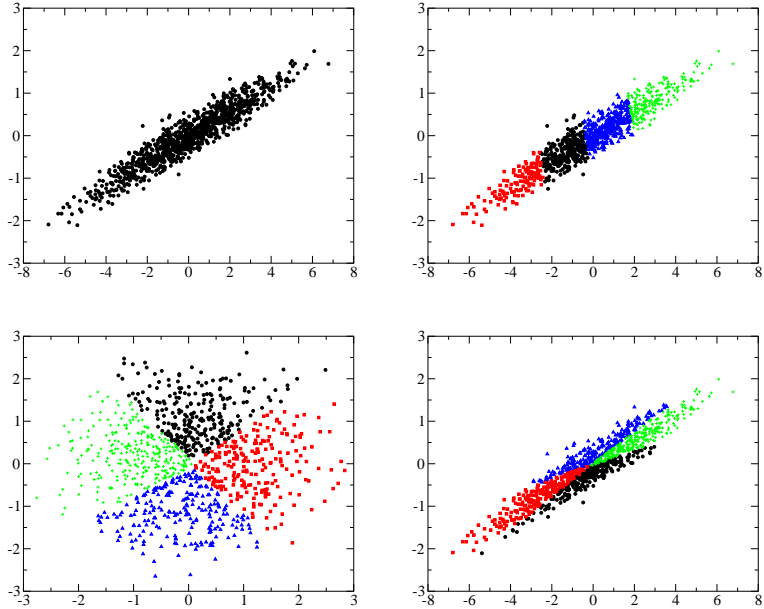}}
      \caption{An example of sphering and K-means clustering using a $2$ parameter
               training set. 
               The top left figure is the original data set. 
               In the top right figure the training set has been clustered into 
               $4$ regions. 
               Since the variation of the parameters is significantly larger in the
               horizontal direction, the clustering simply divides along this axis.
               However, the power spectrum will vary equally in both directions
               of parameter space, so there is a much larger variation in the power 
               spectrum along the vertical direction of each cluster.
               This property can be avoided by sphering the parameter space prior
               to clustering.
               The bottom left figure shows the data set after sphering and then 
               clustering.
               In this basis the parameters and the power spectrum vary equally in all
               directions. 
               The bottom right figure shows the data set back in the original basis. 
               Now there is no bias in the clusters based on the scale or correlations 
               of the parameters. 
               The clusters retain the property that the power spectrum will vary 
               equally across each cluster.}
      \label{clustpics}
   \end{center}
\end{figure*}

In Figure (\ref{clustpics}) we demonstrate the results of using the 
K-means clustering algorithm on a $2$ dimensional parameter space, $\mathcal{N}_{x}=2$.  
The different point types distinguish the members of each of the $4$ clusters. 
Note the difference in the arrangement of the clusters when the data is sphered
prior to clustering.
The sphered data ignores the scale and correlations of the parameters,
giving clusters over which the power spectra vary roughly equally.

\subsection{Power spectrum compression}\label{compression}
\begin{figure}
   \begin{center}
      \resizebox{50mm}{!}{\includegraphics{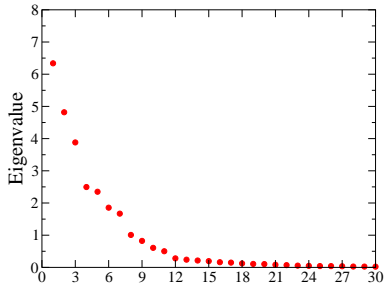}}
      \caption{Plot of the $30$ largest eigenvalues of the covariance matrix 
               of the training set power spectra.  There are $135$ total eigenvalues.
               The fact that only a few of
               the eigenvalues dominate indicates that the power spectra can be
               compressed significantly by rotating into a new basis and projecting
               out directions corresponding to small eigenvalues. 
               In the examples below we keep the directions corresponding to the
               $60$ largest eigenvalues.}
      \label{eigenvals}
   \end{center}
\end{figure}

The efficiency of the algorithm can be improved by using Karhunen-Lo\`{e}ve compression 
\citep{Karhunen:1946,Loeve:1955,Tegmark:1994ed}
to transform the power spectra subspace of the training set to 
a new, lower dimensional space.
We begin with a training set of $10^4$ power spectra generated as in \ref{test1}.
The training set consists of vectors $\mathbf{y}_{j}$
formed from concatenating the scalar TT, TE, and EE power spectra evaluated at
the $45$ ``usual'' $\ell$-values used by CMBfast out to $\ell=1500$.
The ``usual'' $\ell$-values are the ones actually evolved by CMBfast or CAMB;
the power spectrum is interpolated at the intermediary $\ell$'s.
After constructing the covariance matrix of the power spectra $C_{yy}$,
an eigen-decomposition gives a transformation matrix $V$ having the property 
\begin{equation*}
  V C_{yy} V^{T} = E_{yy},
\end{equation*}
where $E_{yy}$ is a diagonal matrix containing the eigenvalues of $C_{yy}$.  
In Figure (\ref{eigenvals}), we have plotted the $30$ largest eigenvalues of $C_{yy}$.
The fact that these eigenvalues vary over a large range indicates
that some redundancy remains in the components of $\mathbf{y}$.  
By choosing a new basis nearly all of the information in the $3$ power spectra 
can be stored in significantly fewer coefficients. 
The compression matrix is formed by dropping the rows of $V$, which are the 
eigenvectors of $C_{yy}$, that have small eigenvalues (relative to the largest).
Then $V$ is a mapping from a $135$ dimensional space to a much smaller 
($\sim 60$ dimensional) space.  
Since a set of polynomial regression coefficients is needed for each 
component of $\mathbf{y}$, this compression algorithm provides a 
significant reduction in the computation time and memory requirements
of the algorithm.

\begin{figure*}
   \begin{center}
      \begin{tabular}{ccc}
      \resizebox{150mm}{!}{\includegraphics{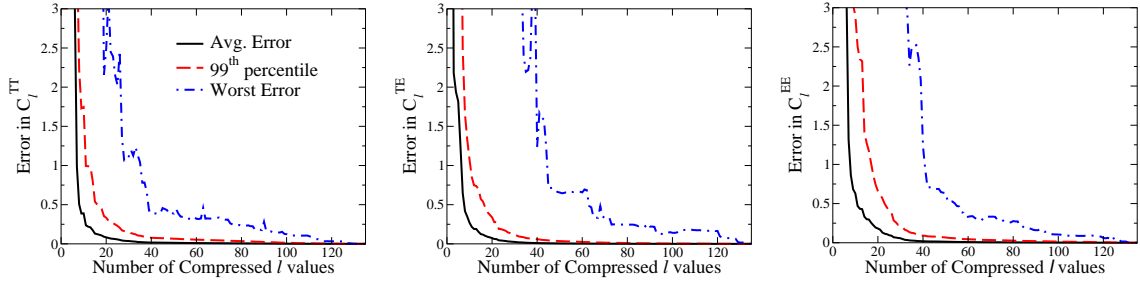}}
      \end{tabular}
      \caption{Plots of the error accrued due to compressing the power spectra 
               as a function of the number of compressed $\ell$'s. The $3$ lines
               denote the average error, the $99$-percentile error bar, and the
               worst error over the $10^{4}$ models in the test set. 
               The worst error is the largest deviation from CAMB at any $\ell$ 
               in any member of the test set.
               For nearly all of the models, only $\sim 60$ compressed $\ell$'s are needed 
               of the $135$ possible to maintain sufficient accuracy in the 
               compressed power spectra.}
    \label{comp}
  \end{center}
\end{figure*}

In Figure (\ref{comp}) we plot the error accrued due to the compression of 
the power spectra.  
That is, we computed $V$, truncated the specified number of rows and calculated
\begin{equation*}
   \mathbf{y}\prime = V^{T} V \mathbf{y}
\end{equation*}
over the $10^4$ models, computed using CAMB, that we will use as our test set 
in section (\ref{results}).
The dimensionless average error in the plots is defined as the mean absolute deviation:
\begin{equation*}\label{err}
   \mbox{Error} = \frac{1}{{\cal N}_{y}} \sum_{\ell} \left< 
       \frac{|C_{\ell} - C_{\ell}^{\mathrm{CAMB}}|}
            {\sigma^{\mathrm{CV}}_{\ell}}
       \right>,
\end{equation*}
where $C_{\ell}$ and $C_{\ell}^{\mathrm{CAMB}}$ denote the individual power
spectra that make up $\mathbf{y}\prime$ and $\mathbf{y}$ respectively.
The brackets denote averaging over the $10^4$ models and 
$\sigma^{\mathrm{CV}}_{\ell}$ is the cosmic standard deviation
computed using $C_{\ell}^{\mathrm{CAMB}}$. 
Recall that the cosmic standard deviation of the TT, TE and EE spectra are given by
\begin{equation*}
   \sigma^{\mathrm{CV, TT}}_{\ell} = \sqrt{\frac{2}{2\ell+1}} C^{TT}_{\ell},
\end{equation*}
\begin{equation*}
   \sigma^{\mathrm{CV, TE}}_{\ell} = \sqrt{\frac{1}{2\ell+1} \left(
                C^{TT}_{\ell} C^{EE}_{\ell} + \left(C^{TE}_{\ell}\right)^{2}\right)},
\end{equation*}
\begin{equation*}
   \sigma^{\mathrm{CV, EE}}_{\ell} = \sqrt{\frac{2}{2\ell+1}} C^{EE}_{\ell}.
\end{equation*}

\end{document}